\newlength{\intwidth}

\documentclass[lineno]{jfm}

\usepackage{amssymb}

\usepackage{array}
\usepackage{amsmath}
\usepackage{latexsym}
\usepackage{bezier}
\usepackage{psfrag,graphicx}
\usepackage{bm}
\usepackage{float}
\usepackage{multirow}
\usepackage{mathrsfs}
\usepackage{color}
\usepackage{enumerate}
\usepackage{natbib}
\usepackage{nicefrac}
\usepackage[colorlinks, citecolor=blue]{hyperref}
\usepackage{overpic}
\usepackage{rotating}
\usepackage{verbatim}

\begin{document}

\title[Multiscale coherent structures in shear flows]{Multiscale quasi-time periodic coherent structures in shear flows}

\author{
  Runjie Song\aff{1},
  Kengo Deguchi\aff{1},
  Genta Kawahara\aff{2}
  \and Yongyun Hwang\aff{3}
}
 
 \affiliation{
   \aff{1}School of Mathematics, Monash University, VIC 3800, Australia
         \aff{2}Graduate School of Engineering Science, University of Osaka, 1-3 Machikaneyama, Toyonaka, Osaka 560-8531, Japan
                  \aff{3}Department of Aeronautics, Imperial College London, London, SW7 2AZ, UK
}

\maketitle

\begin{abstract}
Attempts to disentangle shear-flow turbulence often focus on identifying relatively simple solutions, such as travelling waves or periodic orbits. We show, however, that capturing multiscale features requires considering states at least as complex as quasi-time-periodic solutions. Approximations of these states can be computed efficiently using a quasi-linear model, consistent with the large-Reynolds-number asymptotic analysis. The quasi-linear structure is key to producing multiscale critical layers that generate vortices obeying Taylor’s frozen-flow hypothesis.
\end{abstract}

\section{Introduction}\label{sec:introduction}

Coherent structures in wall-bounded shear flows play a key role in transporting momentum and mass in turbulent flows \citep{townsend1976structure}. 
Understanding and modelling their underlying mechanisms are important for a wide range of engineering applications and environmental sciences, and recent efforts have been driven by two distinct motivations.

A natural motivation is to construct reduced models that faithfully reproduce turbulent direct numerical simulations or experimental data.
A representative approach of this type is quasi-linear (QL) approximations, in which the flow is typically decomposed into slow- and fast-scale components. The governing equations for the fast-scale component are then linearised about the slow-scale flow, with a suitable closure that often neglects the self-interacting nonlinear terms. 
Several research groups have pursued this approach. Recently, \cite{Bretheim_2015} proposed to use a streamwise average to define the slow component and observed reasonable agreement with direct numerical simulation data at moderate Reynolds numbers. 
At higher Reynolds numbers, however, the spectral features of the data produced by the model do not match those of the full Navier-Stokes equations, leading to the development of more complicated models; see \cite{Hernández_Yang_Hwang_2022A}, which also provides a comprehensive historical review of modelling approaches of this type.

Conversely, one can be motivated to extract the minimal mechanisms underlying coherent structures. Even simple solutions, such as travelling-wave or time-periodic solutions known as exact coherent states (ECS, \cite{Waleffe_1998}), were found to be underpinned by the self-sustaining process \citep{hamilton1995}. 
Although these solutions are unstable, dynamical systems theory allows them to be connected to the statistical properties of chaotic states \citep{Kawahara_2012,page2024,W25}.
Attempts to reproduce travelling-wave ECS using reduced models have been made, for example, by \cite{Blackburn_Hall_Sherwin_2013}, \cite{Beaume2015} and \cite{Pausch_2019}, which are closely related to the aforementioned QL models. A careful study of why such models succeed ultimately leads to the large-Reynolds-number asymptotic analysis known as vortex-wave interaction \citep[VWI,][]{hall1991,HALL_SHERWIN_2010}; for a historical overview of the related study, 
see \cite{Deguchi_Hall_2016}. 

In asymptotic analysis, one must define perturbation expansions for the relevant physical quantities and thoroughly verify that they are consistent with the governing equations. 
At present, no reduced model exists that simultaneously satisfies these strict criteria and reproduces turbulence statistics. 
ECS satisfy the governing equations accurately, which is why they are preferred over filtered turbulence data in equation-based approaches such as asymptotic analysis.
Nevertheless, a fundamental dilemma remains: the ECS identified so far struggle to capture some important aspects of turbulence. In particular, multiscale vortical structures and their hierarchical features are absent; see e.g. \cite{Doohan_Bengana_Yang_Willis_Hwang_2022}. 
Our goal is to overcome this limitation by constructing a reduced model that remains consistent with the VWI. The key is that, by incorporating multiple neutral waves into the VWI, each providing feedback to the streaky mean flow, the model can capture more complex temporal behaviour than a travelling wave.

The paper is structured as follows. In \S \ref{sec:2}, we introduce our proposed method, QL-VWI, and highlight the difference from the existing models discussed above. In \S \ref{sec:3}, we first demonstrate that the QL-VWI can approximate a time-periodic Navier-Stokes solutions, and then turn to the more practically relevant plane Poiseuille flow to capture quasi-time-periodic solutions.
The latter solutions capture some, though not all, of the characteristic features of coherent structures in shear flow turbulence, as further discussed in \S \ref{sec:4}.

\section{Formulation of the problem}\label{sec:2}

Consider the Navier-Stokes equations in Cartesian coordinates $(x,y,z)$
\begin{eqnarray}\label{NSeq}
(\partial_t +\mathbf{u}\cdot \nabla)\mathbf{u}=-\nabla p+R^{-1}\nabla^2 \mathbf{u}+R^{-1}Q\mathbf{e}_x,\qquad \nabla \cdot \mathbf{u}=0,
\end{eqnarray}
where $t$ denotes time and $\nabla$ is the gradient operator.
The velocity $\mathbf{u}=(u,v,w)$ and the pressure $p$ are defined in a channel with $y\in [-1,1]$. The combination of the boundary conditions $\mathbf{u}=(\pm 1,0,0)$ at $y=\pm 1$ and $Q=0$ corresponds to plane Couette flow, whose base flow is $\mathbf{u}=(y,0,0)$. Imposing $\mathbf{u}=\mathbf{0}$ at $y=\pm 1$ and $Q=2$ yields plane Poiseuille flow, with base flow $\mathbf{u}=(1-y^2,0,0)$. For both cases, our Reynolds number $R$ is based on the characteristic velocity of the base flow and the channel half width.

We enforce periodicity in $z$ and quasi-periodicity in $x$ and $t$. More specifically, the flow is assumed to depend on $y$, $z$, and the phase variables 
$\theta_m=\alpha_m x+\omega_mt$, $m=1,2,\dots,M$, each taken to be $2\pi$ periodic. Here, $\alpha_m$ and $\omega_m$ are real numbers representing the wavenumbers and frequencies, respectively. 
We use $\beta$ to indicate the $2\pi/\beta$ periodicity of the flow in $z$. Travelling wave, periodic, and quasi-periodic solutions in a standard periodic box can all be captured within this setup.


Now, let us apply a `Reynolds decomposition'  $\mathbf{u}=\overline{\mathbf{u}}+\tilde{\mathbf{u}}$ and $p=\overline{p}+\tilde{p}$, 
defined by averaging over all phase variables. 
Here and hereafter, the overline denotes this averaging operation, while the tilde indicates the fluctuation (or \textit{wave})  component.
The streamwise mean velocity $\overline{u}$ represents the \textit{streaks}, while the cross-stream components $\overline{v}$ and $\overline{w}$ represent the \textit{rolls}. 
Substituting the decomposition into (\ref{NSeq}) and 
applying the averaging yields:
\begin{eqnarray}
[\overline{v}\partial_y +\overline{w}\partial_z-R^{-1}(\partial_y^2+\partial_z^2)]
\left[ \begin{array}{c} \overline{u}\\ \overline{v}\\ \overline{w} \end{array} \right]
+ \left[ \begin{array}{c} -R^{-1}Q\\ \partial_y \overline{p} \\ \partial_z \overline{p}  \end{array} \right] = \mathbf{F},~~~\label{BRE}
\partial_y\overline{v}+\partial_z\overline{w}=0,~~~~~~~\label{BREBRE}
\end{eqnarray}
where $\mathbf{F}=-\overline{(\tilde{\mathbf{u}}\cdot {\nabla})\tilde{\mathbf{u}}}$ is the Reynolds stress.
Taking the difference between (\ref{NSeq}) and (\ref{BRE}) gives the fluctuation part of the momentum equations 
\begin{eqnarray}
(\partial_t+\overline{\mathbf{u}}\cdot {\nabla})\tilde{\mathbf{u}}+
(\tilde{\mathbf{u}}\cdot {\nabla})\overline{\mathbf{u}}
+\{(\tilde{\mathbf{u}}\cdot {\nabla})\tilde{\mathbf{u}}-\overline{(\tilde{\mathbf{u}}\cdot {\nabla})\tilde{\mathbf{u}}}\}
=-{\nabla} \tilde{p}+R^{-1}{\nabla}^2 \tilde{\mathbf{u}},~~~\label{waveeq2}
\end{eqnarray}
and ${\nabla}\cdot \tilde{\mathbf{u}}=0$. At this stage, no approximations have been made to the equations.

The terms in curly brackets in (\ref{waveeq2}) 
are the wave-wave interactions, which we hereafter refer to as WW.
Removing them makes (\ref{waveeq2}) linear in the fluctuations. 
Using the normal mode expansions 
$\tilde{\mathbf{u}}=\sum_{m=1}^M\hat{\mathbf{u}}_m(y,z)\exp(\text{i}\theta_m)+\text{c.c.}$ and $\tilde{p}=\sum_{m=1}^M \hat{p}_m(y,z)\exp(\text{i}\theta_m)+\text{c.c.}$, with c.c. indicating the complex conjugate, the fluctuation part of the governing equations reduces to
\begin{subequations}\label{waveeq}
\begin{eqnarray}
\mathscr{L}\hat{u}_m+\hat{v}_m\partial_y \overline{u}+\hat{w}_m\partial_z \overline{u}=-\text{i}\alpha_m \hat{p}_m,\\
\mathscr{L}\hat{v}_m+\hat{v}_m\partial_y \overline{v}+\hat{w}_m\partial_z \overline{v}=-\partial_y \hat{p}_m,\\
\mathscr{L}\hat{w}_m+\hat{v}_m\partial_y \overline{w}+\hat{w}_m\partial_z \overline{w}=-\partial_z \hat{p}_m,\\
\text{i}\alpha_m \hat{u}_m+\partial_y\hat{v}_m+\partial_z\hat{w}_m=0,
\end{eqnarray}
\end{subequations}
where $\mathscr{L}=\text{i}\alpha_m(\overline{u}-c_m)+\overline{v}\partial_y+\overline{w}\partial_z-R^{-1}(\partial_y^2+\partial_z^2-\alpha_m^2)$. The phase speeds $c_m=-\omega_m/\alpha_m$ are real, so the waves are neutral.
At large $R$, the behaviour of the above equations is well approximated by the Rayleigh equation, which has a singularity at the critical level where $\overline{u}=c_m$ (see \cite{HALL_SHERWIN_2010}).

Among the terms in the wave equation (\ref{waveeq}), we refer to those involving the roll components $\overline{v},\overline{w}$ as the WR terms. Also, the streamwise component of the Reynolds stress 
\begin{eqnarray}\label{Fwave}
\mathbf{F}=
-\sum_{m=1}^M\left[ \begin{array}{c} \partial_y(\hat{u}_m\hat{v}_m^*)+\partial_z(\hat{u}_m\hat{w}_m^*)\\ \partial_y(\hat{v}_m\hat{v}_m^*)+\partial_z(\hat{v}_m\hat{w}_m^*)\\ \partial_y(\hat{w}_m\hat{v}_m^*)+\partial_z(\hat{w}_m\hat{w}_m^*)\end{array} \right]+\text{c.c.},
\end{eqnarray}
is referred to as the SR term. 
The QL-VWI approach solves (\ref{waveeq}) and (\ref{BREBRE}) with the right hand side replaced by (\ref{Fwave}). In \cite{Beaume2015}, \cite{Deguchi_Hall_2016}, and \cite{Pausch_2019}, the same system was solved but with $M=1$; the 
approach of \cite{Blackburn_Hall_Sherwin_2013} is a further simplification in which the WR and SR terms are neglected. The common feature between the above models and the QL model mentioned in \S 1 is that the WW terms are omitted. However, while the former assumes the waves to be neutral and seeks approximations to ECS, the latter allows for the temporal evolution of the waves directly through numerical integration, so that the solution to the fluctuation equations corresponds to neutral Floquet/Lyapunov mode (for periodic/chaotic case).  There are numerous variants of the QL type model; for example, \cite{Thomas_2014} considered multiple waves within a certain wavenumber band, whereas \cite{Alizard_Biau_2019} used only a single wave component.
In the generalised QL approach, some of the WW terms are retained \citep[see][]{Hernández_Yang_Hwang_2022A}.

Asymptotic analysis of the quasi-periodic solutions of the Navier-Stokes equations naturally leads to a multiple-wave extension of VWI. This conclusion remains unchanged even when the governing equations are replaced by QL-VWI, meaning that our model retains all of the dominant terms associated with VWI-type coherent structures at large Reynolds numbers.
The asymptotic expansion has leading-order scalings $\overline{u}=O(R^0), \overline{v}=O(R^{-1}), \overline{w}=O(R^{-1})$, $\hat{\mathbf{u}}_m=O(\delta^{1/2}R^{-1})=O(R^{-7/6})$, except within the critical layers of thickness $\delta=R^{-1/3}$.
These layers emerge around the critical levels to regularise the singularity of the Rayleigh equation.
Within its corresponding critical layer, the $m$th wave is amplified as $\hat{\mathbf{u}}_m=O(\delta^{-1/2}R^{-1})=O(R^{-5/6})$. The streamwise vorticity is most pronounced within the critical layer and is therefore a suitable quantity for visualising the waves.
The Reynolds stress due to the waves is concentrated within the critical layer and influences the outer roll flow through stress jump conditions across the critical level (see \cite{HALL_SHERWIN_2010} for further details). These jump conditions are difficult to impose numerically; therefore,  \cite{Blackburn_Hall_Sherwin_2013} derived their model using viscous regularisation of the VWI system.

It should be emphasised that, after substituting the asymptotic expansion, the leading-order system is determined solely by the assumption of large $R$. At this stage, we cannot freely choose which terms to keep or drop, unlike in other modelling approaches. This order-of-magnitude comparison confirms that the WW, WR, and SR terms are indeed negligible. 
%
We also remark that  
the VWI theory can be extended so that the overlined components evolve on a slow time scale while the tilde components evolve on a fast time scale \cite[see][]{HALL_SHERWIN_2010}. 
However, the simultaneous temporal evolution of roll-streaks and waves makes the separation between fast and slow scales unclear. The asymptotic analysis is therefore justified only for ECS; yet their properties can indirectly influence the more complex surrounding dynamics, as suggested by dynamical systems theory (see \cite{page2024,W25}).

The results in the next section are obtained by solving the QL-VWI system using the Newton-Raphson method. The parameters $R$, $\alpha_m$, and $\beta$ are prescribed, whereas $c_m$ are determined as part of the solution. We use a Chebyshev-Fourier spectral method for discretisation, which allows us to adapt the fully nonlinear Navier-Stokes solver of \cite{Deguchi_Hall_Walton_2013} 
with only minor modifications. To fully resolve the sharp critical layer structure, we typically employ up to 300 Chebyshev polynomials in the $y$-direction and up to 50 Fourier harmonics in the $z$-direction. At the selected points, we increased these numbers to 400 and 60, respectively, and confirmed that the results are well converged.

\section{Numerical results} \label{sec:3}

\S \ref{sec:PCF} shows that the parameter range of plane Couette flow studied in \cite{Deguchi_Hall_2016} serves as an ideal test ground of the QL-VWI approach. The key observation is that the gentle periodic orbit (GPO) solution found by \cite{KAWAHARA_KIDA_2001} exhibits a standing-wave-like structure, realised through two counter-propagating symmetric waves. 
\S \ref{sec:PPF} aims to compute QL-VWI approximation of quasi-periodic states in plane Poiseuille flow. Such quasi-periodic solutions are difficult to obtain directly from the full Navier–Stokes equations, yet they are essential for forming multiscale structures.

\subsection{GPO in plane Couette flow}\label{sec:PCF}

\begin{figure}
\centering
\begin{overpic}[width=0.9 \textwidth]{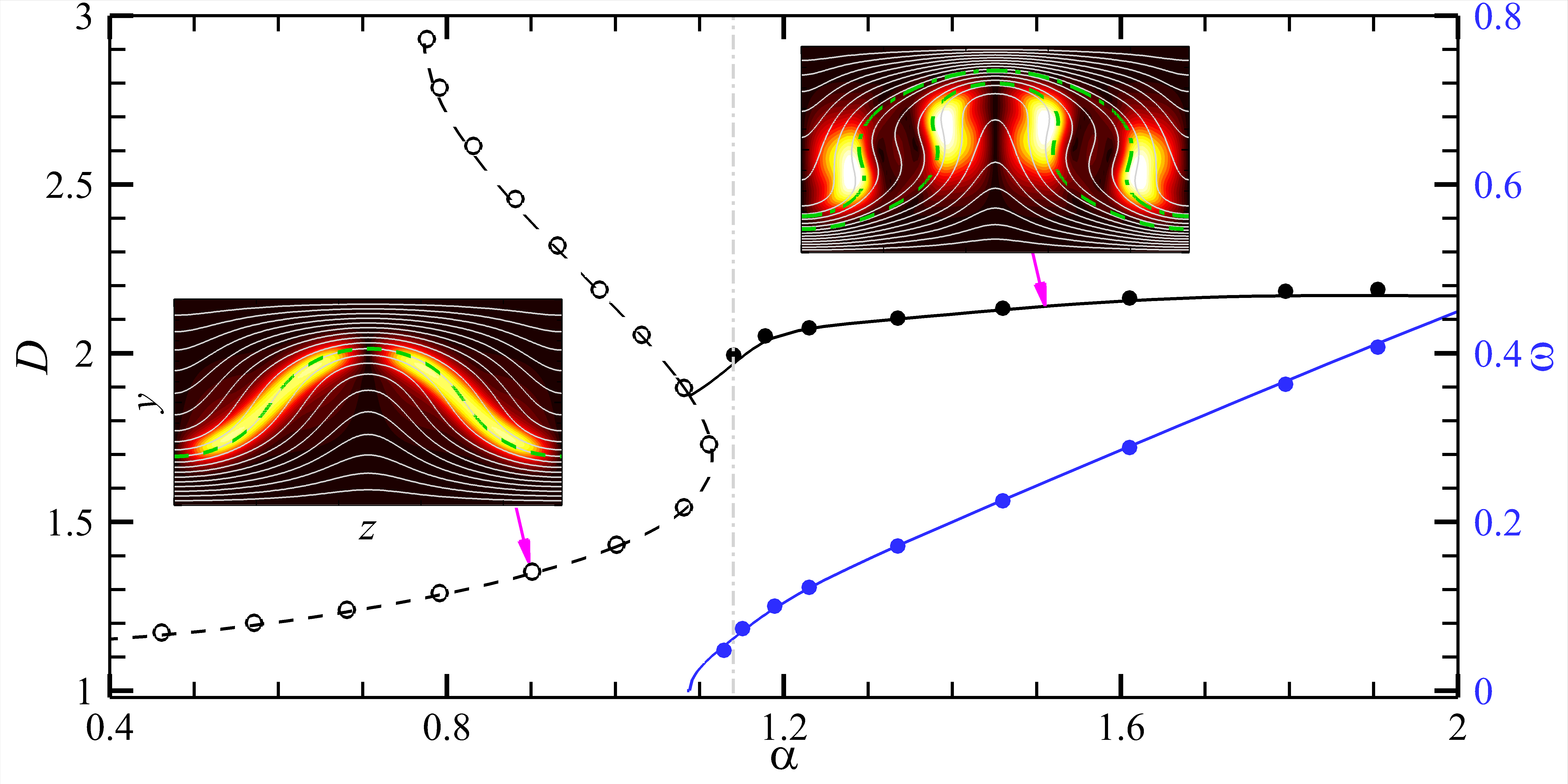}
\end{overpic}
\caption{Comparison of Navier–Stokes results (symbols) and QL-VWI results (lines) in plane Couette flow at $R=3000$. Our choice of $\beta=1.667$ ensures that, at $\alpha=1.14$ (indicated by the vertical line), the solutions fit the minimal box used in \cite{hamilton1995}. The black plots show the drag on the lower wall $D$, i.e. the $x$--$z$--$t$ average of $\partial_y u$ at $y=-1$. Dashed lines and open circles represent the NBW, while solid lines and filled circles represent the GPO. The frequency $\omega$ of the GPO is shown by the blue plots. The left and right colourmaps show $\tilde{u}_{\text{rms}}$ of the NBW ($\alpha=1$) and the GPO ($\alpha=1.5$), respectively, computed using the QL-VWI. The plotted domain spans $y\in [-1,1]$ and $z\in [0,2\pi/\beta]$. The green lines denote the critical levels. The grey contours are the level curves of $\overline{u}$. 
}
\label{fig:fig1}
\end{figure}

Figure~\ref{fig:fig1} examines how varying the streamwise period of the periodic box $[0,\frac{2\pi}{\alpha}]\times [-1,1]\times [0,\frac{2\pi}{\beta}]$  affects the ECS at $R=3000$. 
The black open circles show the drag on the wall $D$ of the steady solution of Navier-Stokes equations known as NBW \citep{Nagata_1990,Clever_Busse_1992,Waleffe_1998}. When $\alpha$ is increased (i.e. the streamwise period of the computational box is shortened)
above that of the minimal box \citep{hamilton1995}, the NBW disappears at a turning point and is replaced by the GPO  (filled circles). 

As reported by \cite{Deguchi_Hall_2016}, the NBW result is well approximated by the QL-VWI with $M=1$, $\alpha_1=\alpha$ (black dashed line). The black solid line in the figure illustrates our novel result: the QL-VWI system with $M=2$, $\alpha_1=\alpha_2=\alpha$, provides an excellent approximation of the GPO drag.
This solution branch is obtained via bifurcation analysis of the NBW.
The two waves in the approximated GPO exhibit a symmetry, satisfying $c_1=-c_2$. The blue open circles show the frequency $\omega=|c_1\alpha_1|$ of the QL-VWI solution. 
This agrees well with  $\omega=2\pi/T$ computed from the period $T$ of the full Navier-Stokes GPO, providing unequivocal validation of our approximation. 
We repeated a similar QL-VWI computation at $R=20000$ and found that, for $\alpha<1.7$, the result at $R=3000$ is already asymptotically converged. Computing the full Navier-Stokes GPO at $R=20000$ is challenging, but we observed excellent agreement with the QL-VWI solution at several representative values of $\alpha$ (e.g. for $\alpha=1.25$, the error in the approximation of $D$ and $\omega$ is less than 0.2\%).

The colourmaps of figure~\ref{fig:fig1} show the root-mean-square of the streamwise wave component,
$\tilde{u}_{\text{rms}}(y,z)=(2\sum_{m=1}^M|\hat{u}_m|^2)^{1/2}$. 
As the left panel shows, the NBW is supported by a steady wave that is concentrated around the critical level where the streak $\overline{u}$ vanishes (green dashed line).
On the other hand, as shown in the right panel, the GPO has two critical levels at $\overline{u}=c_1=-0.162$ (green dashed line) and $\overline{u}=c_2=0.162$ (green dot-dashed line). At the bifurcation point, the two critical levels merge into one, and the GPO coincides with the NBW. 

\begin{figure}
\centering
\begin{overpic}[width=0.8 \textwidth]{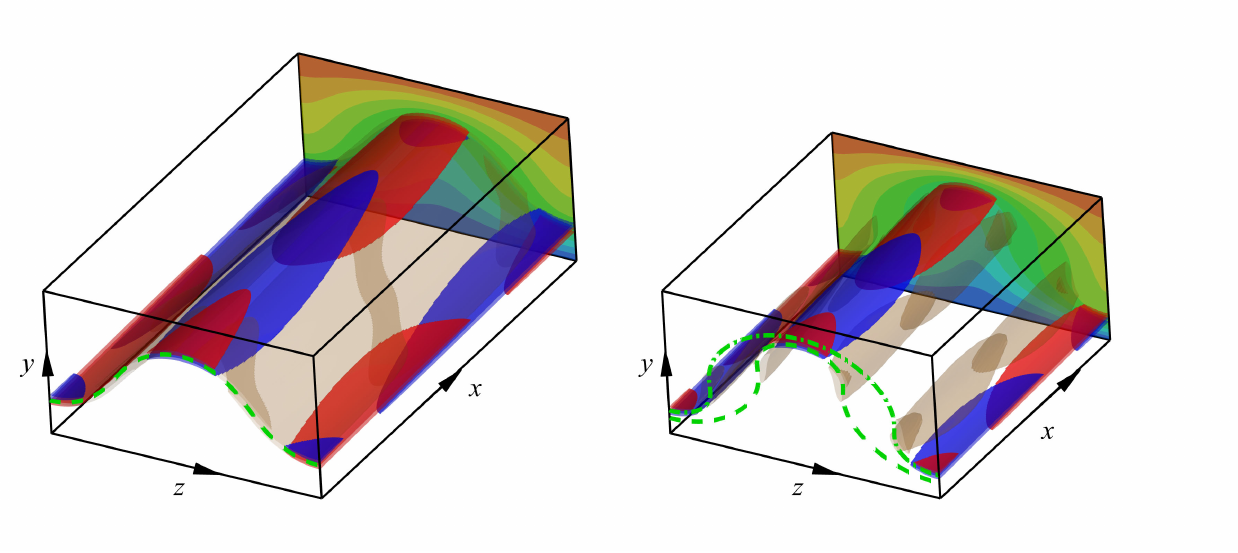}
\put(2,42){(a)}
\put(50,42){(b)}
\end{overpic}
\caption{Flow visualisation of the NBW (a) and the GPO (b) at $R=20000$ obtained by the QL-VWI.
The colourmaps denote the streamwise velocity of the streak field and the blue/green isosurfaces denote the 50\% maximum/minimum
value of $\partial_y\tilde{w}-\partial_z \tilde{v}$. The yellow isofurfaces show the 50\% maximum value of $|\tilde{u}|$. 
}
\label{fig:fig2}
\end{figure}
As $R$ increases, the wave becomes increasingly concentrated near the critical level, consistent with the VWI theory. Figure~\ref{fig:fig2} shows three-dimensional plots of the QL-VWI solutions obtained at $R=20000$. The yellow transparent surfaces represent the magnitude of the streamwise velocity, $|\tilde{u}|$, while the red and blue surfaces show the streamwise vorticity, $\partial_y\tilde{w}-\partial_z \tilde{v}$. The Navier-Stokes results are almost indistinguishable from those shown in figure \ref{fig:fig2} and are therefore omitted.

\subsection{Multiscale solution in plane Poiseuille flow}\label{sec:PPF}

\begin{figure}
\centering
\begin{overpic}[width=0.95 \textwidth]{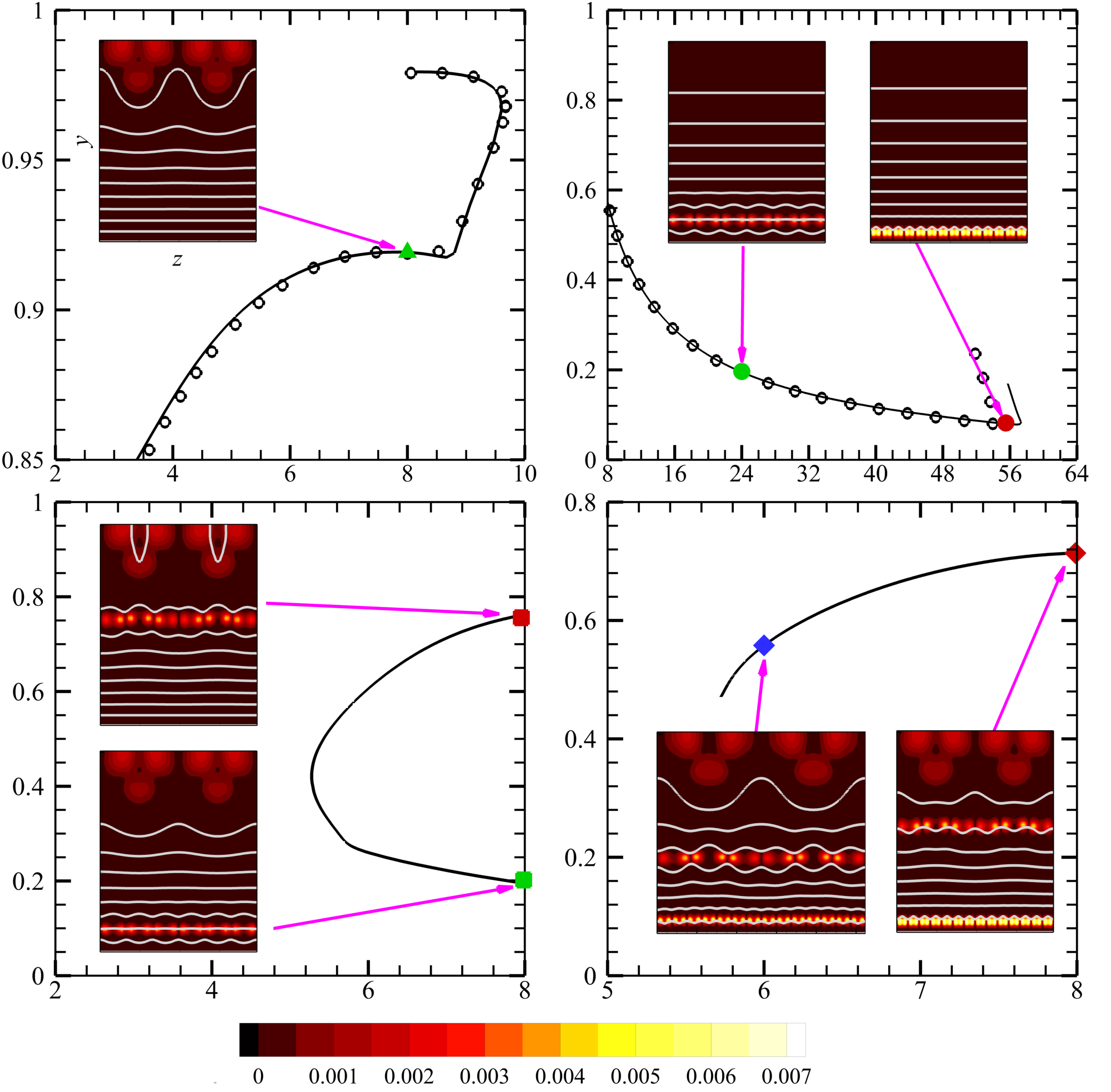}
\put(-2,52){(c)}
\put(47.5,52){(d)}
\put(-2,31){$c_2$}
\put(-2,96){(a)}
\put(-2,76){$c$}
\put(25,7.5){$\beta$}
\put(75,7.5){$\beta$}
\put(47.5,96){(b)}
\end{overpic}
\caption{
Bifurcation diagram of plane Poiseuille flow solutions. Lines show the QL-VWI results and  symbols are the Navier-Stokes results.
The colourmaps use the same format as figure 1; all panels show the domain $(y,z) \in [-1,0]\times [0,2\pi/8]$, except for one.
(a) TWA with $\alpha/\beta=3/4$.
(b) TWB with $\alpha/\beta=1/2$. 
(c) $N=2$ QL-VWI computation with $\alpha_1/\beta=6/8$ and $\alpha_2/\beta=12/8$.
(d) $N=3$ QL-VWI solution with $\alpha_1/\beta=6/8$,  $\alpha_2/\beta=12/8$ and $\alpha_3/\beta=28/8$.
The colourmap of the solution indicated by the blue diamond shows the domain $(y,z) \in [-1,0]\times [0,2\pi/6]$. 
}
\label{fig:fig4_DNS}
\end{figure}
Our starting point consists of two Navier-Stokes travelling-wave solutions. The first is the solution independently discovered by \cite{Gibson_Brand_2014} and \cite{Deguchi_2015}, which we call TWA in this paper. The second is the solution recently found by \cite{Song_Deguchi_2025}, referred to as MS-A3, which we denote TWB here for simplicity. Throughout this section, we set $R=10^5$.

Figure~\ref{fig:fig4_DNS}-(a) shows the bifurcation diagram of TWA, obtained by varying $\beta$ while fixing the streamwise-spanwise aspect ratio of the wave, $\alpha/\beta$. 
The symbols show the phase speed $c$ of the travelling waves, and the solid lines are 
the result obtained using the $M=1$ version of the QL-VWI system (write $c_1=c$, $\alpha_1=\alpha$ hereafter). As observed by \cite{Deguchi_2015}, TWA in the high-wavenumber regime exhibits coherent structures localised near the centre of the channel. The colourmap in panel (a) shows the same type of visualisation as in figure~\ref{fig:fig1}, restricted to the lower half of the channel.
By symmetry of the solution, a pair of streaks appears within one 
$z$-period.
Figure~\ref{fig:fig4_DNS}-(b) shows analogous computational results for TWB. We find that, in the high-wavenumber range, this solution localises near the wall, as illustrated by the colourmap.

Note that the colourmaps in panels (a) and (b) are plotted over the same domain, $[-1,0]\times[0,2\pi/8]$. TWA at the green triangle corresponds to $\beta=8$ and $\alpha=(3/4)\beta=6$, while TWB at the green circle has $\beta=24$ and $\alpha=(1/2)\beta=12$. Thus, the colourmap for the latter solution includes 3 copies within the domain.

The next step is to construct a new QL-VWI solution by superposing these two travelling wave solutions.
To reduce computational cost, it is desirable that the target solution possesses symmetry about the channel centreline; therefore, we also include a $y$-symmetric copy of TWB, denoted TWBr, in the superposition. In the QL-VWI with $N=2$, we set $\beta=8$, $\alpha_1=6$, and $\alpha_2=12$, and make the following substitutions:
\begin{enumerate}
\item ~~assign the fluctuation part of TWA to $\hat{\mathbf{u}}_1, \hat{p}_1$,
\item ~~assign the sum of the fluctuation parts of TWB and TWBr to $\hat{\mathbf{u}}_2, \hat{p}_2$,
\item ~~assign the sum of the mean parts of TWA, TWB and TWBr to $\overline{\mathbf{u}}, \overline{p}$,
\item ~~assign the phase speeds of TWA and TWB to $c_1$ and $c_2$, respectively.
\end{enumerate}
After the substitution, the residual of the equations is small but nonzero, as the system is nonlinear. We treat this residual as a forcing term and gradually reduce its amplitude as a continuation parameter in the Newton-Raphson tracking of the solution branch (i.e., a homotopy approach). The resulting solution is the green square in figure~\ref{fig:fig4_DNS}-(c). Note that the substitution of TWB and TWBr actually uses three repeated copies of each solution so that they fit into the $\beta=8$ computational domain. This repetitiveness is, of course, slightly broken after the homotopy.

The solution branch in figure~\ref{fig:fig4_DNS}-(c) is obtained by fixing $\alpha_1/\beta$ and $\alpha_2/\beta$. 
After the turning point, 
the critical layer of the second wave moves toward the channel centre. This creates space near the wall, allowing us to insert the localised solution at the red circle in figure~\ref{fig:fig4_DNS}-(b) ($(\alpha,\beta)=(28,56)$, so its colourmap contains 7 repeated copies). The superposition of the red square solution, the red circle solution, and its symmetric counterpart can be carried out in QL-VWI with $N=3$. 
By choosing $(\alpha_1,\alpha_2,\alpha_3)=(6,12,28)$ and applying the homotopy method, we obtain the red diamond solution shown in figure~\ref{fig:fig4_DNS}-(d). 

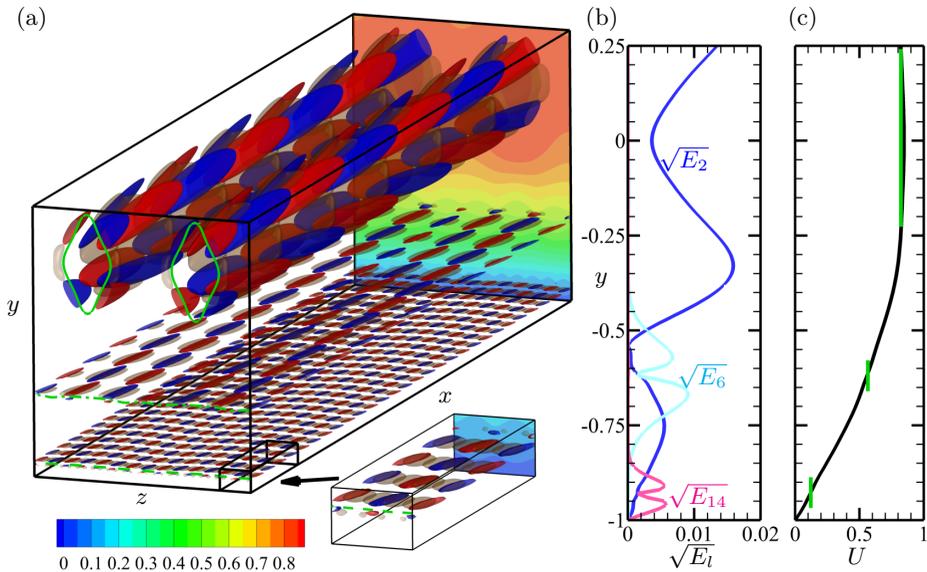
\begin{figure}
\centering
\begin{overpic}[width=0.9 \textwidth]{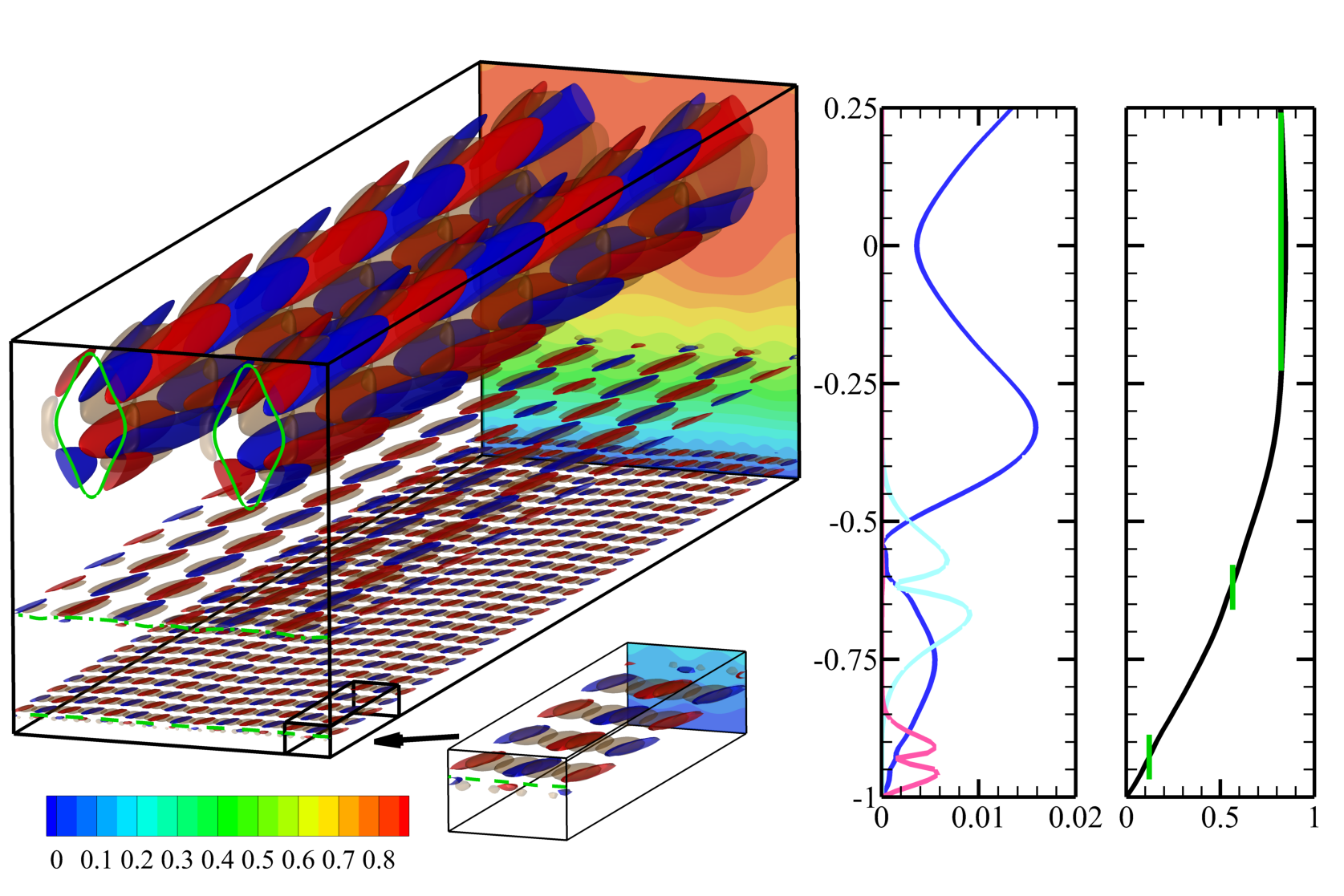}
\put(-1,61){(a)}
\put(45,20){$x$}
\put(-2,30){$y$}
\put(12,9){$z$}
\put(61,61){(b)}
\put(83,61){(c)}
\put(89.5,2.5){$U$}
\put(69,45.5){\textcolor{blue}{$\sqrt{E_2}$}}
\put(71,22){\textcolor{cyan}{$\sqrt{E_6}$}}
\put(70,9){\textcolor{magenta}{$\sqrt{E_{14}}$}}
\put(62,33){$y$}
\put(70,2.5){$\sqrt{E_l}$}
\end{overpic}
\caption{The flow field of the solution at the blue diamond in figure~\ref{fig:fig4_DNS}-(d).
(a) Visualisation in the box $[0,2\pi/1.5]\times [-1,0.25] \times[0,2\pi/6]$. See figure~\ref{fig:fig2} caption for definitions of the surfaces and colourmap. (b) Energy of the streak associated with the $l$th spanwise Fourier mode.
(c) The $x$--$z$--$t$ mean streamwise velocity $U(y)$. The green vertical lines indicate the projected locations of the critical layers, whose horizontal positions correspond to the associated phase speeds. 
}
\label{fig:fig5}
\end{figure}
Note that the QL-VWI approximation breaks down at the Kolmogorov scale (see \cite{Deguchi_2015} for details). Indeed, a close inspection of figure~\ref{fig:fig4_DNS}-(b) shows that the red circle lies close to the scale at which the approximation begins to fail. Therefore, for safety, in figure~\ref{fig:fig4_DNS}-(d) we vary $\beta$ from 8 to 6 fixing $\alpha_1/\beta,\alpha_2/\beta$ and $\alpha_3/\beta$. 
Inspecting the flow field of the solution marked by the blue diamond (i.e. $(\alpha_1,\alpha_2,\alpha_3)=(4.5,9,21)$, the streamwise period becomes $2\pi/1.5$), we find that this final adjustment has the side benefit of producing nearly equispaced critical layers.
Figure~\ref{fig:fig5}-(a) shows a three-dimensional visualisation of this solution in the periodic domain, plotted in the same format as figure~\ref{fig:fig2}. Its implications for shear-flow turbulence are discussed in the next section.

\section{Discussion} \label{sec:4}

We have constructed approximations to ECS using the QL-VWI model, a quasi-linear formulation consistent with the multi-wave VWI theory.
The plane Couette flow results in \S~\ref{sec:PCF} show that QL-VWI reproduces GPO with remarkably high accuracy, even at moderate $R$. This suggests that the QL-VWI solution for plane Poiseuille flow shown in figure~\ref{fig:fig5} also provides a good approximation to the corresponding quasi-periodic Navier-Stokes solution. 
The latter is difficult to obtain using a standard Navier-Stokes approach, whereas the QL-VWI requires only a computational cost comparable to that for a steady state.

Two important features of shear flow turbulence can be seen in the solution shown in figure~\ref{fig:fig5}. First, the vortices exhibit a hierarchy in which their size decreases towards the wall, echoing Townsend's attached eddy hypothesis \citep{townsend1976structure}. 
This is evident from panel (a), where the isosurfaces and colourmap represent the characteristic structures of the wave and the streak, respectively. Interestingly, observations of turbulent shear flows by \cite{Hwang2015} revealed that small-scale vortices in the log layer are not attached to the wall, whereas streamwise-elongated structures are. The latter can be interpreted as the streak component in VWI theory (although the roll-streaks of our solution are steady, the theory can be extended to allow these structures to depend on slow spatiotemporal scales; see e.g.  \cite{Song_Deguchi_2025}). 
Figure~\ref{fig:fig5}-(b) shows the distribution of energy contained in the $l$th spanwise Fourier mode of the streak, $E_l$. The largest-scale mode, represented by $E_2$, suggests that large-scale streaks are attached to the wall. While smaller scale modes, $E_6$ and $E_{14}$ exhibit almost self-similarity when scaled by their respective spanwise wavelengths, suggesting that the assumption typically used to describe log layer dynamics is realised.

Second, the solution satisfies Taylor’s frozen-flow hypothesis \citep{Taylor_1938}; that is, eddies are convected by the local mean velocity. This is demonstrated in figure~\ref{fig:fig5}-(c), where the location of the critical layer and the phase speed are compared with the mean flow $U(y)$. 
As seen by comparing panels (a) and (c), probes inserted at $y\approx -0.9,-0.6$ and $-0.1$ should record wave signals with different characteristic frequencies; such features cannot appear in previously known ECS.
Note however that critical-layer structures tend to be blurred in turbulent flows due to the non-negligible effects of the WW term (see e.g. \cite{Hernández_Yang_Hwang_2022A}). The WW term is the origin of the turbulent cascade, and our solution correspond to an idealised setting in which such complexity is absent. 
Moreover, constructing ECS whose mean flow more closely resembles turbulence would require the inclusion of many more waves in the computation. While this task is straightforward in principle, we leave it for future work.

In summary, 
our study narrows the gap between ECS and fully developed Navier-Stokes turbulence.
Recalling that the two features discussed above are typically introduced as assumptions in turbulence theories, it is intriguing that they can instead be rationally justified on the basis of high-Reynolds-number asymptotic analysis for our ECS. Extending our idea, identifying ECS that are even closer to turbulence and analysing their structure in a systematic manner could provide a valuable complement to data-driven studies.

\backsection[Acknowledgements]{
This research was supported by the Australian Research Council Discovery Project DP230102188. }

\backsection[Declaration of Interests]{
The authors report no conflict of interest.
}

\bibliographystyle{jfm}  
\bibliography{Reference}  
\end{document}